\documentclass{article}
\usepackage{graphicx}
\usepackage{amsmath}

\input{tcilatex}

\begin{document}

\begin{center}
{\LARGE Anomalous Single Production of the Fourth SM Family Quarks and
Leptons at Future Electron-Positron Colliders}

\bigskip

Y. Islamzade$^{1)}$,{\large \ }H. Karadeniz$^{1,2)}$, S. Sultansoy$^{1,3)}$
\end{center}

$^{1)}$Dept. of Physics, Faculty of Arts and Sciences, Gazi University,
06500 Teknikokullar, Ankara, Turkey.

$^{2)}$Ankara Nuclear Research and Training Center, 06100 Besevler, Ankara,
Turkey.

$^{3)}$Institute of Physics, Academy of Sciences, H. Cavid Ave. 33, Baku,
Azerbaijan.

\bigskip

\bigskip

\begin{center}
\textbf{Abstract}
\end{center}

We study the possibility to produce single fourth SM family fermions at
electron-positron colliders via anomalous $\gamma -f_{4}-f$ interactions.

\bigskip

\textbf{I. Introduction}

\bigskip

Flavor Democracy [1-4] forces [5-7] the existence of the fourth fermion
family in the Standard Model (SM). In this case, the smallness of the first
three neutrino masses gains a natural explanation [8] without see-saw
mechanism. In addition these small masses are compatible\ with large mixing
angles, assuming that neutrinos are of the Dirac type [9]. The latest
precision electroweak data allows the existence of the extra SM\ families
[10,11]. Moreover, as quoted from [11] ''quality of the fit for the one new
generation is as good as for zero new generations (Standard Model)''.\
Current experimental lower limits on the fourth SM family fermions masses
are: 92.4 GeV for $l_{4}$, 45.0 (83.3) GeV for stable (unstable) Dirac $\nu
_{4}$, 199 (128) GeV for d$_{4}$ decaying via neutral-current
(charged-current) channel $\left[ 12\right] .$ \ 

The fourth family quarks will be copiously produced at the LHC [12,13].
These quarks lead to an assential enhancement of the Higgs boson production
via gluon-gluon fusion at the Tevatron and LHC and, as a consequence, the
''golden mode'' (H$\longrightarrow ZZ\longrightarrow 4$l) becomes important
even at the upgraded Tevatron [14,15]. On the other hand, future lepton
colliders will be advantageous for investigation of the fourth SM family
leptons and quarkonia [16,17].

Main parameters of proposed e$^{+}$e$^{-}$\ colliders are given in the Table
1. \ As can be seen, the first stage will cover 500 GeV center-of-mass
energy. On the other hand, expected masses of the fourth SM family fermions
are close to each other and lie between 300 and 700 GeV (see [7] and
references therein).\ \ \ For this reason their pair production at the first
stage linear colliders is not promising. Therefore, in this case we must
look for the single production via anomalous interactions. Arguments
presented in [18] concerning t-quarks anomalous interactions are already
valid for fourth family fermions.

In this paper we study single production of the fourth family leptons and
quarks at e$^{+}$e$^{-}$\ colliders. For the time being we restrict
ourselves to the anomalous interactions mediated by photon.

\bigskip

\textbf{II. Anomalous }$\gamma -f_{4}-f$\textbf{\ interactions \bigskip }

Within the Standard Model there are no anomalous vertices like $\gamma
-f_{4}-f$ at the tree level, they arise only due to loop contributions and,
therefore, are suppressed. On the other hand, these vertices are essentially
enhanced in various extensions of the SM. For example, the possibility of
anomalous transitions exist in a wide class of dynamical models for the
fermion mass generation [18]. Following [19,20], the $\gamma -f_{4}-f$\ \
vertices can be written as 
\begin{equation}
\Gamma _{\mu }^{\gamma }\text{=k}_{\gamma }\frac{\text{ee}_{f}}{\Lambda }%
\sigma _{\mu \nu }\left( \text{g}_{1}\text{P}_{l}+\text{g}_{2}\text{P}_{%
\text{r}}\right) \text{q}^{\nu }  \tag{1}
\end{equation}
where $\Lambda $ is the new physics cutoff, $\ e=\sqrt{4\pi \alpha }$ , $%
e_{f}=$-1, 2/3 and -1/3 for $l_{4},u_{4}$ and $d_{4},$ respectively, k$%
_{\gamma }$ is the strength of the anomalous coupling, $\sigma ^{\mu \nu }$= 
$\left( \gamma ^{\mu }\gamma ^{\nu }-\gamma ^{\nu }\gamma ^{\mu }\right) /2$%
, P$_{l}$ =$\left( 1+\gamma _{5}\right) /2$, P$_{r}=\left( 1-\gamma
_{5}\right) /2$, g$_{1}$and g$_{2}$ denote the relative magnitude of the
left and right components of the $f_{4}-f$ current, $g_{1}^{2}+g_{2}^{2}=1.$
Parametrization used in [18] is obtained from (1) by replacing. 
\begin{equation*}
\frac{C_{\gamma }}{m_{t}}\text{=k}_{\gamma }\frac{\text{ee}_{f}}{\Lambda }%
\frac{g_{1}+g_{2}}{2}\text{, \ }\frac{D_{\gamma }}{m_{t}}\text{=k}_{\gamma }%
\frac{\text{ee}_{f}}{\Lambda }\frac{g_{1}-g_{2}}{2}
\end{equation*}

\bigskip

\textbf{III. Decays of the fourth family fermions}

\bigskip

Decay width for the SM modes is given by, 
\begin{equation*}
\Gamma \left( f_{4}\longrightarrow f^{\prime }W\right) =\frac{G_{F}}{8\pi 
\sqrt{2}}\left| V_{4i}\right| ^{2}M^{3}\sqrt{1-\frac{(m_{W}+m_{f})^{2}}{M^{2}%
}}\sqrt{1-\frac{(m_{W}-m_{f})^{2}}{M^{2}}}
\end{equation*}

\begin{equation}
\left[ 1+\frac{(m_{W}+m_{f})^{2}}{M^{2}}+\frac{m_{W}m_{f}}{M^{2}}-2\frac{%
(m_{W}^{2}-m_{f}^{2})^{2}}{M^{4}}\right]  \tag{2}
\end{equation}
where V$_{4i}$ is the corresponding element of the 4x4 CKM matrix, M denotes
the mass of the fourth family fermion. Anomalous interactions lead to
additional FCNC decay modes. Let us remind that, at this stage we constrain
ourselves with anomalous interactions mediated by photon only. According to
(1) we obtain: 
\begin{equation}
\Gamma \left( f_{4}\longrightarrow f\gamma \right) =\text{k}_{\gamma }^{2}%
\frac{\alpha \text{e}_{f}^{2}}{4}\frac{M^{3}}{\Lambda ^{2}}(1-\frac{m_{f}^{2}%
}{M^{2}})\left( 1+\frac{m_{f}^{2}}{M^{2}}-2\frac{m_{f}^{4}}{M^{4}}\right) 
\tag{3}
\end{equation}
As seen from (2) SM decay modes are determined by CKM matrix elements.
Within the parametrization given in [6-8] the main SM decay modes are u$%
_{4}\longrightarrow $bW$^{+},$\newline
$d_{4}\longrightarrow $tW$^{-},$ $l_{4}\longrightarrow \nu _{\tau }$W$^{-},$ 
$\nu _{4}\longrightarrow \tau $W$^{+}$ and corresponding CKM matrix elements
are $\left| V_{u_{4}b}\right| \simeq \left| V_{d_{4}t}\right| \simeq 0.005,$ 
$\left| V_{l_{4}\nu _{\tau }}\right| \simeq \left| V_{\nu _{4}\tau }\right|
\simeq 10^{-4}$, respectively.

Differing from the case of t-quark decays $\left[ 19\right] $, where SM
decay mode is dominant with respect to decay modes mediated by anomalous
interaction, in decays of fourth family fermions anomalous interaction mode
may be well dominating. Neglecting masses of W-boson and first three family
fermions compared to fourth family fermion masses, which are taken to be
equal to 400 GeV for numerical estimations, and assuming the cutoff scale $%
\Lambda $ to be equal to m$_{4}$ we get from (2) and (3) following rough
values of k$_{\gamma }^{2}$ which indicates dominance of anomalous mode: $%
\left| k_{\gamma }\right| ^{2}>$ 2.5$\times $10$^{-7}$ for $l_{4}$, $\left|
k_{\gamma }\right| ^{2}>$ 1.4$\times $10$^{-3}$ for u$_{4}$ and $\left|
k_{\gamma }\right| ^{2}>$ 5.6$\times $10$^{-3}$ for d$_{4}$.

\bigskip

\textbf{IV. Total cross section for single production of fourth family
fermions}

\bigskip

Using Eq. (1) for anomalous $\gamma -f_{4}-f$ \ vertex one can easily obtain
the cross section for the process e$^{+}$e$^{-}\rightarrow \gamma ^{\ast
}\rightarrow $f$_{4}\overline{\text{f}}$, 
\begin{equation*}
\sigma =\frac{N_{c}}{3}\frac{\pi \alpha ^{2}}{s}k_{\gamma }^{2}\frac{M^{2}}{%
\Lambda ^{2}}e_{f}^{2}\frac{s}{M^{2}}\sqrt{1-\frac{(M+m_{f})^{2}}{s}}\sqrt{1-%
\frac{(M-m_{f})^{2}}{s}}
\end{equation*}
\begin{equation}
\left[ 1+\frac{(M+m_{f})^{2}}{s}+\frac{Mm_{f}}{s}-2\frac{%
(M^{2}-m_{f}^{2})^{2}}{s^{2}}\right]  \tag{4}
\end{equation}

\bigskip

Cross section values as a function of the fourth family fermions mass for $%
\sqrt{s}$ = 500 GeV are presented in the Fig. I. Taking 25 events per
working year (10$^{7}$ s) as the observation limit, we see that $\sqrt{s}$ =
500 GeV e$^{+}$e$^{-}$ colliders with integrated luminosity of 100 fb$^{-1}$%
can reach the following upper limits for $\left| k_{\gamma }\right| ^{2}$:
3.6$\times $10$^{-3}$ for $\overline{u}_{4}u(c),$ 1.66$\times $10$^{-2}$ for 
$\overline{d}_{4}d(s)$ and 5.4$\times $10$^{-3}$ for $\overline{l}_{4}e(\mu
,\tau ).$ With M = 400 GeV the process e$^{+}$e$^{-}\rightarrow \gamma
^{\ast }\rightarrow \overline{\text{u}}_{4}$t is kinematically forbidden at $%
\sqrt{s}$ = 500 GeV.

\bigskip

\textbf{V. Conclusion}\bigskip

Although the first stage (500 GeV c.m. energy) of future lineer
electron-positron colliders is not promising for the pair production of the
fourth SM family fermions, it will yield important results for the single
production via anomalous interaction. As an example, we show that
sensitivity up to 0.01 can be reached for $\left| k_{\gamma }\right| ^{2}$.

We studied single production via anomalous interaction with photon only. The
production via Z is possible as well. In this case both $k_{\gamma }$ and $%
k_{Z}$ must be taken into account together. The corresponding analysis is
being studied and will be reported in the next work.

\bigskip

\textbf{Acknowledgements}

\bigskip

This work is supported in part by Turkish State Plannig Organization under
the Grant No DPT 2002K120250.

\bigskip

\textbf{\bigskip\ References}

[1] H. Harari, H. Hant and J. Weyers, Phys. Lett. B 78 (1978) 171.

[2] H. Fritzsch, Phys. Lett. B 184(1987) 391.

[3] H. Fritzsch and J. Plankl, Phys. Lett. B 237 (1990) 451.

[4] H. Fritzsch and D. Holtmannspotter, Phys. Lett. B 388 (1994) 290.

[5] A. Datta and S. Raychaudhuri, Phys. Rev. D 49 (1994) 4762.

[6] A. \c{C}elikel, A. K. \c{C}ift\c{c}i and S. Sultansoy; Phys. Lett. B 342
(1995) 257.

[7] S. Sultansoy, Why the Four SM Families, hep-ex/0004271 (2000).

[8] S. Ata\u{g} et al., Phys. Rev. D 54 (1996) 5745.

[9] J. I. Silva-Marcos, Phys. Rev. D 59 (1999) 091301. \ 

[10] H.-J. He, N. Polonsky and S. Su, Phys. Rev. D 64 (2001) 053004.

[11] V. A. Novikov, L. B. Okun, A. N. Rozanov and M. I. Vysotsky, Phys.
Lett. B 529 (2002) 111.

[12] Review of Particle Physics, D. E. Groom et al., Eur. Phys. J. C 15
(2000) 1.

[13] E. Arik et al., Phys. Rev. D 58 (1998) 117701.

[14] ATLAS Collaboration, ATLAS TDR, CERN/LHCC-99-15 (1999).

[15] O. \c{C}akir and S. Sultansoy, Phys. Rev. D 65 (2001) 013009.

[16] E. Arik, O. \c{C}akir, S. A. \c{C}etin and S. Sultansoy, hep-ph/0203257
(2002); to be published in Phys. Rev. D.

[17] A. K. \c{C}ift\c{c}i, R. \c{C}ift\c{c}i and S. Sultansoy, Phys. Rev. D
65 (2002) 013009.

[18] R. \c{C}ift\c{c}i, A. K. \c{C}ift\c{c}i and S. Sultansoy,
hep-ph/0203083 (2002).

[19] H. Fritzsch and D. Holtmannspotter, Phys. Lett. B 457 (1999) 186.

[20] V. F. Obraztsov, S. R. Slabospitsky and O. P. Yushchenko, Phys. Lett. B
426 (1998) 393.

[21] R. D. Peccei and X. Zhang, Nucl. Phys. B 337 (1990) 269.

\bigskip

\begin{center}
\textbf{Table 1. }Parameters of e$^{+}$e$^{-}$ colliders.

\bigskip 
\begin{tabular}{|c|cc|cc|ccc|}
\hline
& TESLA &  & JLC/ & NLC &  & CLIC &  \\ \hline
& 1 & \multicolumn{1}{|c|}{2} & 1 & \multicolumn{1}{|c|}{2} & 1 & 
\multicolumn{1}{|c}{2} & \multicolumn{1}{|c|}{3} \\ \hline
$\sqrt{s}$, GeV & 500 & \multicolumn{1}{|c|}{800} & 500 & 
\multicolumn{1}{|c|}{1000} & 500 & \multicolumn{1}{|c}{1000} & 
\multicolumn{1}{|c|}{3000} \\ \hline
L, 10$^{34}$ cm$^{-2}$s$^{-1}$ & 3 & \multicolumn{1}{|c|}{5} & 2.5 & 
\multicolumn{1}{|c|}{2.5} & 1.4 & \multicolumn{1}{|c}{2.7} & 
\multicolumn{1}{|c|}{10.0} \\ \hline
\end{tabular}

\bigskip

\FRAME{ftbpFU}{3.6789in}{2.5789in}{0pt}{\Qcb{\textbf{\ }The cross section as
a function of M. Dashed, dotted, dashed-dotted and solid curves are
correspond to $\overline{u}_{4}u(c)$, $\overline{u}_{4}t$, $\overline{l}%
_{4}e(\protect\mu ,\protect\tau )$ and $\overline{d}_{4}d(s,b)$,
respectively. Cross section is in fb, while M is in GeV.}}{\Qlb{fig1}}{%
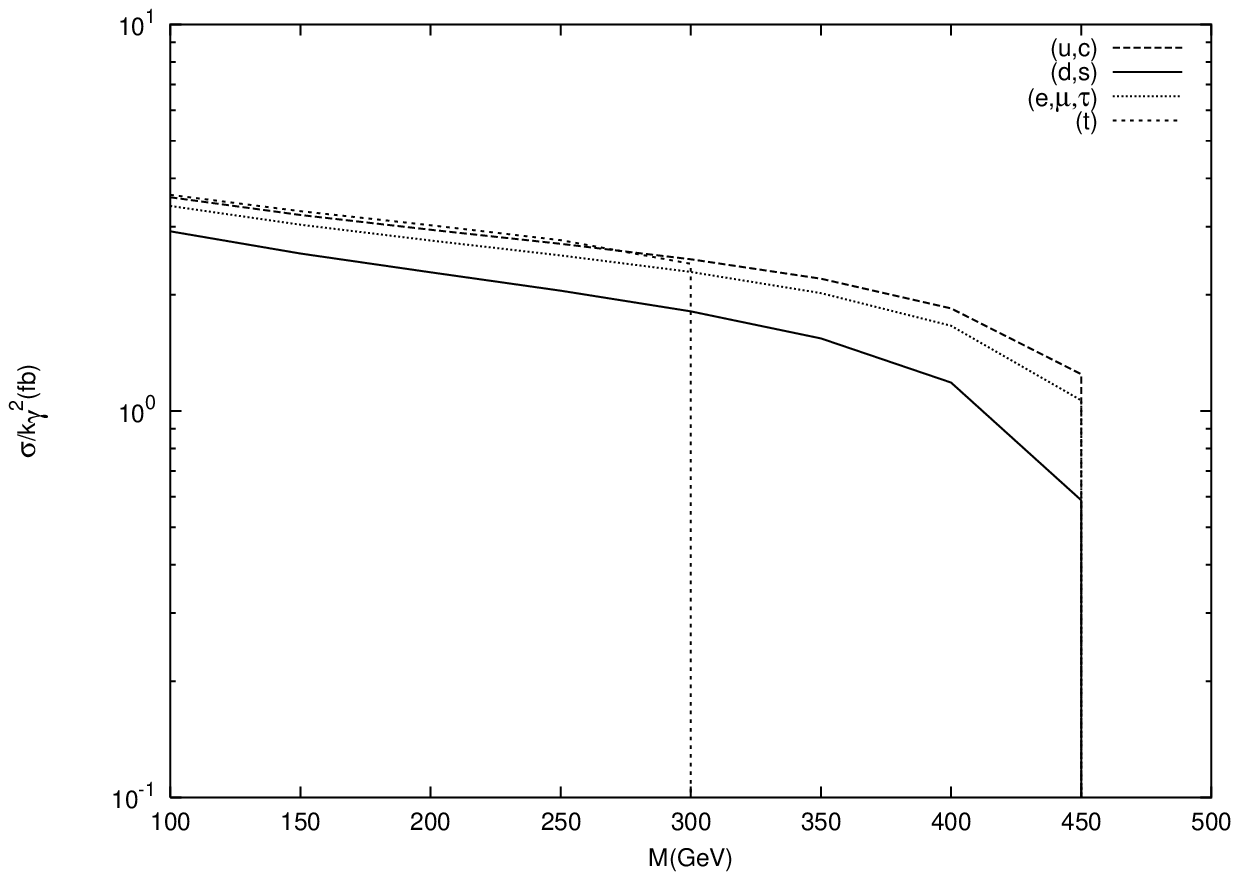}{\special{language "Scientific Word";type
"GRAPHIC";maintain-aspect-ratio TRUE;display "USEDEF";valid_file "F";width
3.6789in;height 2.5789in;depth 0pt;original-width 5.0029in;original-height
3.4964in;cropleft "0";croptop "1";cropright "1";cropbottom "0";filename
'fig1.eps';file-properties "XNPEU";}}
\end{center}

\end{document}